# SPARQL Assist Language-Neutral Query Composer

Luke McCarthy[1], Ben Vandervalk[1], Mark Wilkinson[1]

[1]Providence Heart + Lung Institute at St. Paul's Hospital,
University of British Columbia, Vancouver, BC, Canada
markw@illuminae.com

**Abstract.** SPARQL query composition is difficult for the lay-person or even the experienced bioinformatician in cases where the data model is unfamiliar. Established best-practices and internationalization concerns dictate that semantic web ontologies should use terms with opaque identifiers, further complicating the task. We present SPARQL Assist: a web application that addresses these issues by providing context-sensitive type-ahead completion to existing web forms. Ontological terms are suggested using their labels and descriptions, leveraging existing XML support for internationalization and language-neutrality.

## 1 Introduction

The health care and life science sectors have been some of the most enthusiastic adopters of semantic web technologies. The benefits of the RDF/OWL data model are well-understood by bioinformaticians who have too long had to deal with the problem of integrating data from multiple sources with wildly different underlying schema. These benefits are less obvious, however, to clinicians and researchers who merely see one mysterious query language (SQL) exchanged for another (SPARQL). Even a semantic web-savvy informatician can be daunted when faced with the challenge of querying an unfamiliar data source whose particular RDF vocabulary is initially unknown.

The issue is compounded by the growing use of opaque, semantic-free URIs for ontological classes and properties (OBO [1], SIO [2], CWA [3]). Where the meaning of rdf:type or dc:title in a SPARQL query is relatively clear, the meaning of, for example, sio:SIO_000253 is considerably harder to glean without looking up its ontological definition. Nevertheless, there are many valid reasons for designing ontologies this way, not the least of which is language neutrality.

RDF/XML provides built-in language neutrality by way of the xml:lang attribute; an ontology can easily be internationalized by providing multiple rdfs:label or rdfs:comment properties with appropriate xml:lang attributes. However, even those projects who have, in principle, adopted language neutrality for their classes (e.g. OBO), have not done so for their properties (OBO Relationship Ontology [4]). This is no-doubt due, at least in part, to the difficulty of composing SPARQL queries in which predicates have opaque identifiers. Nevertheless, it is crucial that we do not allow convenience to direct the development of a core global resource - the Semantic Web - and thus the problem should be solved at the level of the tools provided, rather than the resources themselves.

## 2 SPARQL Assist

In this demonstration, we present SPARQL Assist: a web application that facilitates the construction of SPARQL queries by providing context-sensitive type-ahead completion. In addition to assistance with basic syntax, ontological terms are indexed by their labels, allowing a query to be composed in a user's preferred language, assuming appropriate labels are present in the ontology. Terms are read on-the-fly from any ontology specified in a FROM clause, but SPARQL Assist can also be configured to pre-load terms from particular ontologies or SPARQL endpoints.

The entire query, as it is being constructed, is used to provide context for the type-ahead suggestions. Previously declared variables or known individuals are suggested in the subject or object position of a clause and known properties are suggested in the predicate position. If a clause specifies an individual, properties that individual is known to have are displayed preferentially. Similarly, if a clause contains a variable that can ultimately be connected to a known individual in another part of the query, that connection is used to find the most likely properties in the current clause.

Terms are cached on the client side to speed up repeated look-ups, but most of the processing is done on the server side in Java to take advantage of the mature OWL toolkit on that platform. In the future, as much computation as possible will be transferred to the client side to improve both performance and flexibility of deployment.

For this demonstration, SPARQL Assist has been implemented in the context of creating queries that will be resolved by the Semantic Health and Research Environment (SHARE [5]).

## 3 SHARE

SHARE is an advanced SPARQL query client built on top of the SADI Framework [6] for Semantic Web Services. In SADI, services attach properties to input OWL instances and are indexed in a central registry based on the properties they attach. SHARE maps the triple patterns of a SPARQL query onto these indexed properties, allowing a user to query the entire virtual graph of registered SADI services. The RDF data required to answer a given query is thus dynamically generated in response to that query.

In the context of this demonstration, this infrastructure makes SPARQL query composition even more difficult, since there is no pre-existing database to inspect for candidate properties and individuals. The specialized SPARQL Assist provider for SHARE, therefore, uses the SADI registry, in addition to any loaded ontologies, to suggest properties to be used in a query. As in the generic case, if a clause contains a named individual or a variable previously connected to an individual, that information is used to further refine the suggestions; in this case by filtering services (and the resulting suggested properties) that cannot accept a particular individual.

## 4  Conclusion

SPARQL Assist provides prototype solutions for two important problems. First, to hasten the uptake of Semantic Web technologies, it is important to improve access to, and usability of, Semantic Web resources for the lay-end-user while still maintaining best-practices in the way these resources are modeled. Opaque identifiers for both classes and properties are important, as they allow us to avoid "churn" as an ontology evolves over time. We must therefore support the end-user in constructing queries over resources formatted in this way. Second, the Semantic Web is intended to be a global resource, of use to all. As such, a respect for internationalization is also critical, even at these early stages in Semantic Web evolution. We believe that SPARQL Assist provides motivation to more widely adopt what are clearly best-practices in Semantic Web data provision.

**Acknowledgments.** This work has been supported by the Heart + Stroke Foundation of BC and Yukon, Microsoft Research, The Canadian Institutes for Health Research, The Natural Sciences and Engineering Research Council of Canada, and CANARIE.